\documentclass[runningheads]{llncs}
\usepackage[T1]{fontenc}
\usepackage{graphicx}
\usepackage{graphicx}

\usepackage{caption}
\usepackage{multirow}
\usepackage{times}
\usepackage{amssymb}

\usepackage{amsfonts}
\usepackage{tabularx}
\usepackage{makecell}
\usepackage{mathrsfs}
\usepackage{booktabs} % For formal tables
\usepackage{color}
\usepackage{hyperref} 
\usepackage{url}
\usepackage{multirow}
\usepackage{pifont}
\usepackage{bm}
\usepackage{amsmath}

\usepackage{comment}
\usepackage{wasysym}
\usepackage{threeparttable}
\usepackage[figuresright]{rotating}
\usepackage[utf8]{inputenc}

\usepackage[usenames,dvipsnames]{xcolor}
\usepackage{tikz} 

\usepackage{indentfirst}
\usetikzlibrary{calc, arrows.meta, intersections, patterns, positioning, shapes.misc, fadings, through,decorations.pathreplacing}

\def\1{{\bf{1}}}
\def\0{{\bf{0}}}

\def\i{{\bf i}}
\def\j{{\bf j}}

\def\x{{\bf x}}

\begin{document}
\title{A Comprehensive Survey on Enterprise Financial Risk Analysis from Big Data and LLMs Perspective}
\titlerunning{Enterprise Financial Risk Analysis from Big Data Perspective}
\author{Huaming Du\inst{1} \and
Cancan Feng\inst{2} \and
Yuqian Lei\inst{3} \and
Chenyang Zhang\inst{4} \and
Guisong Liu\inst{1} \and
Gang Kou\inst{1,5,6} \and
Carl Yang\inst{7} \and
Yu Zhao\inst{1}} 
% %
\authorrunning{H. Du et al.}
\institute{Southwestern University of Finance and Economics, Chengdu, China \\\email{dhmfcc@swufe.edu.cn, gliu@swufe.edu.cn, zhaoyu@swufe.edu.cn}\and
Chengdu University, Chengdu, China \and
Universität Hamburg, Hamburg-Eimsbüttel, Germany \and
The University of Hong Kong, Hong Kong, China \and 
Hunan University of Technology and Business, Hunan, China \and Xiangjiang Laboratory, China \and
Emory University, Atlanta, Georgia, United States\\
\email{j.carlyang@emory.edu}}
\maketitle              % typeset the header of the contribution
\begin{abstract}
Enterprise financial risk analysis aims at predicting the future financial risk of enterprises.
Due to its wide and significant application, enterprise financial risk analysis has always been the core research topic in the fields of Finance and Management. Based on advanced computer science and artificial intelligence technologies, enterprise risk analysis research is experiencing rapid developments and making significant progress.
Therefore, it is both necessary and challenging to comprehensively review the relevant studies.
Although there are already some valuable and impressive surveys on enterprise risk analysis from the perspective of Finance and Management, these surveys introduce approaches in a relatively isolated way and lack recent advances in enterprise financial risk analysis. In contrast, this paper attempts to provide a systematic literature survey of enterprise risk analysis approaches from the perspective of Big Data and large language models.
Specifically, this survey connects and systematizes existing research on enterprise financial risk, offering a holistic synthesis of research methods and key insights. We first introduce the problem formulation of enterprise financial risk in terms of risk types, granularity, intelligence levels, and evaluation metrics, and summarize representative studies accordingly. We then compare the analytical methods used to model enterprise financial risk and highlight the most influential research contributions. Finally, we identify the limitations of current research and propose five promising directions for future investigation.
\end{abstract}

\section{Introduction}
Enterprise financial risk analysis has long been a focus of scholarly research in the fields of Finance and Management \cite{billio2012econometric}.
Whether young start-ups, small and medium-sized enterprises (SMEs) or famous Fortune 500 enterprises, to some extent they all inevitably face one or multiple financial risks, such as credit risk, guarantee risk, supply chain risk, bankruptcy risk \cite{Altman1968Financial}. In particular, enterprises face big risk challenges when an economic crisis or a gray rhino incident happened, such as the financial crisis in 2008  and COVID-19 epidemic in 2020.
In the modern economic system, as the global financial system becomes more deeply interconnected, enterprises are exposed to both internal risks and contagion risks \cite{Zhao2022Bankruptcy}.
% , each enterprise will be affected by its own risk or the risk brought by surrounding enterprises. 
Therefore, predicting the financial risk of enterprises is of great importance for both government policymakers and financial institutions.

% Studies on enterprise financial risk analysis, which originated in the fields of Finance and Management, have gradually attracted increasing numbers of researchers from Computer Science (CS). 

The earliest research mainly focused on the three most commonly used statistical econometric methods, e.g. multivariate discriminant analysis, linear probability model, and logistic regression, and studied their application in enterprise risk prediction. With technological advancements in artificial intelligence (AI) and large language models (LLMs), researchers have begun to evaluate enterprise financial risk from a Big Data perspective. Beyond traditional enterprise financial index, more comprehensive enterprise risk intelligence, including non-financial textual information and relational data, is now considered.
Specifically, to deal with textual risk information, NLP techniques, including sentiment analysis and event extraction are used to dig enterprise risk signals from non-financial textual data. To model the risk momentum spillover on enterprise relational data, AI techniques, including deep learning and Graph Neural Networks (GNNs), are utilized to evaluate the enterprise contagion risk. 
% \cite{} \TBD
This series of studies has opened up new avenues for analyzing the generation and contagion mechanisms of enterprise financial risk from a Big Data perspective.

\begin{figure*}[t]
    \centering
    \begin{minipage}{0.47\linewidth}
        \centering
        \includegraphics[width=1\textwidth]{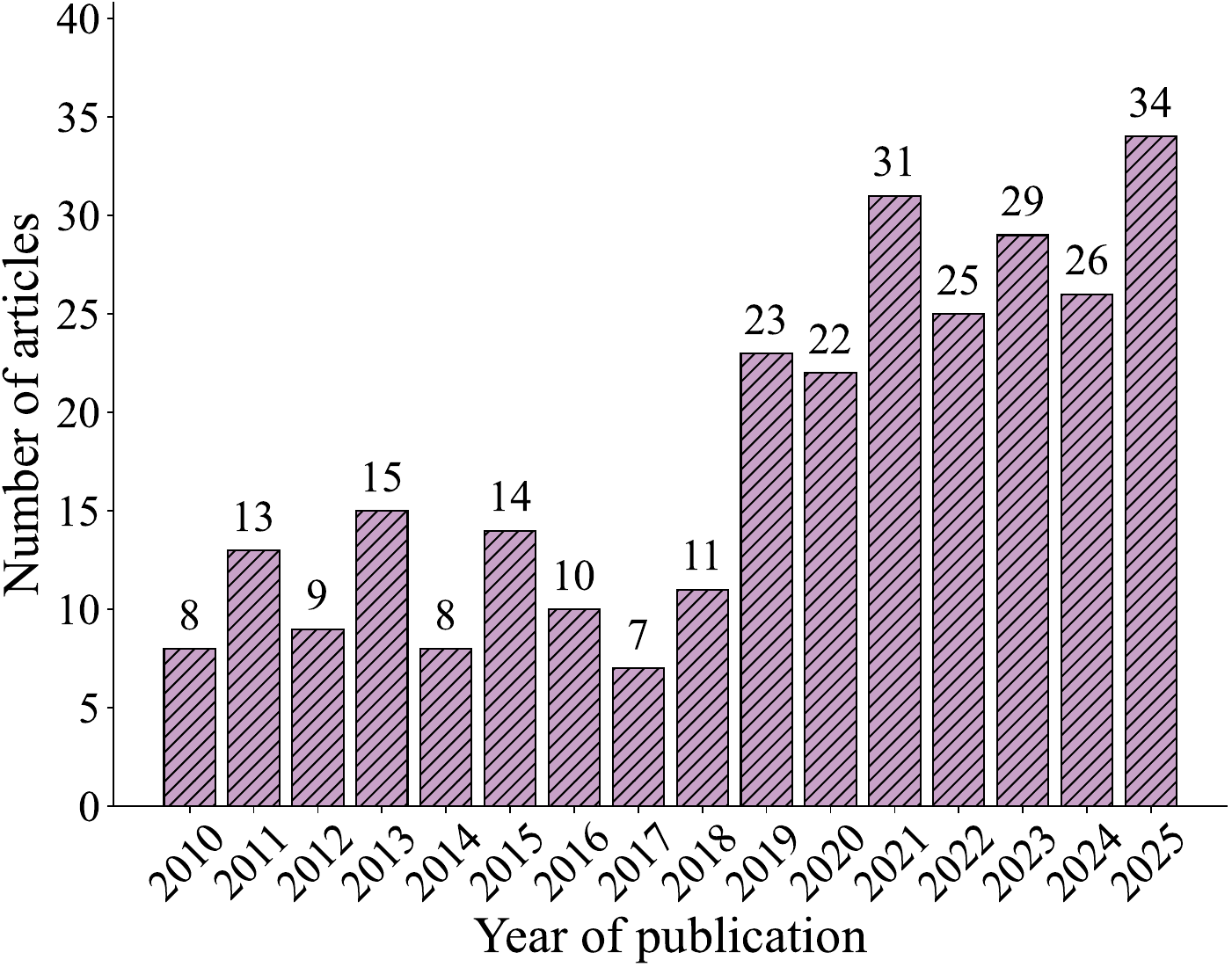}
        \caption{{Publications overview.}}  
        \label{figure-publication-overview}
    \end{minipage}
    \begin{minipage}{0.51\linewidth}
        \centering
        \includegraphics[width=1\textwidth]{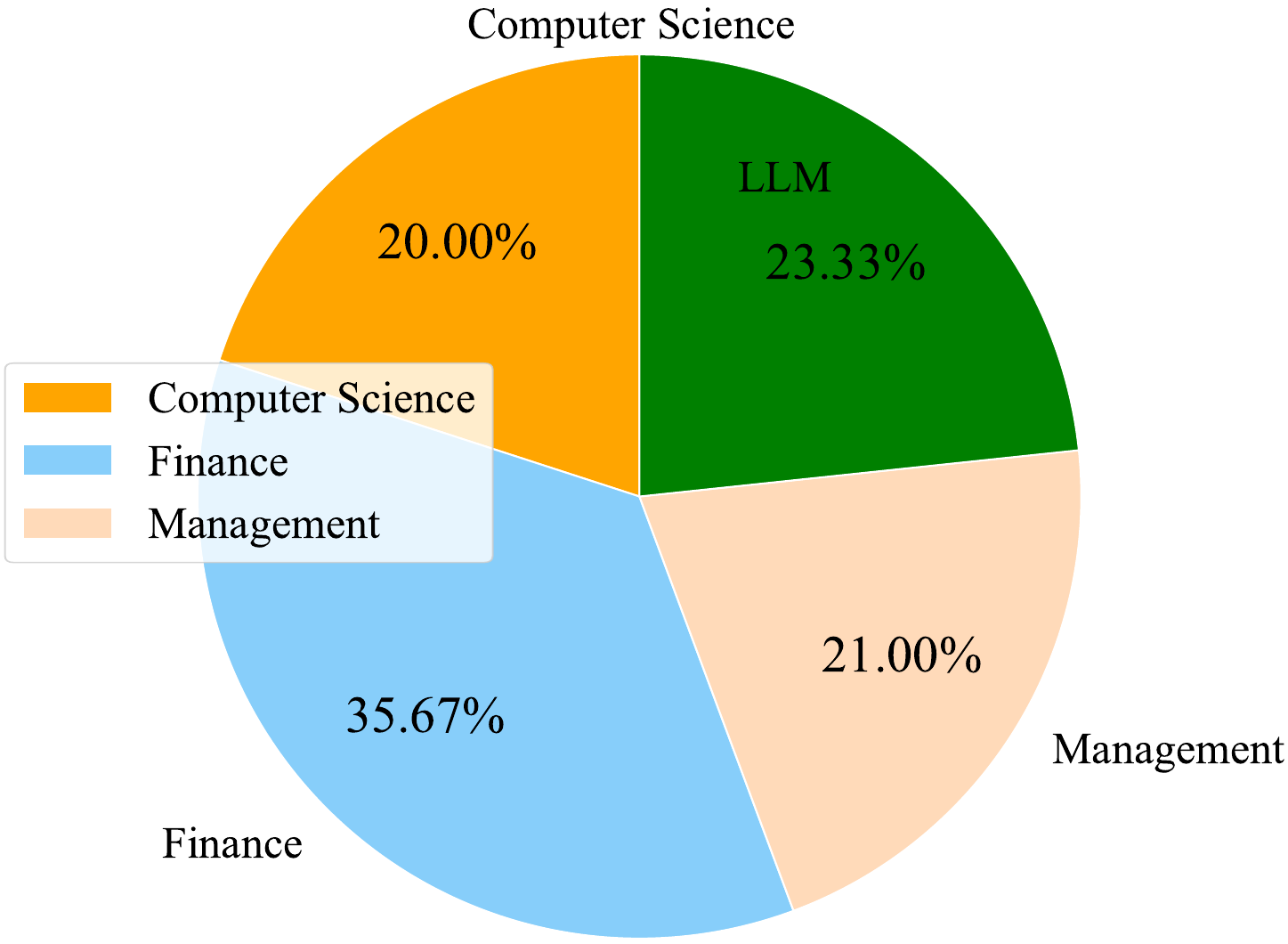}
        \caption{Ratio of research areas.}  
        \label{figure-ration}
        \end{minipage}
\end{figure*}

\begin{figure*}[t]
    \centering
    \includegraphics[width=0.96\textwidth]{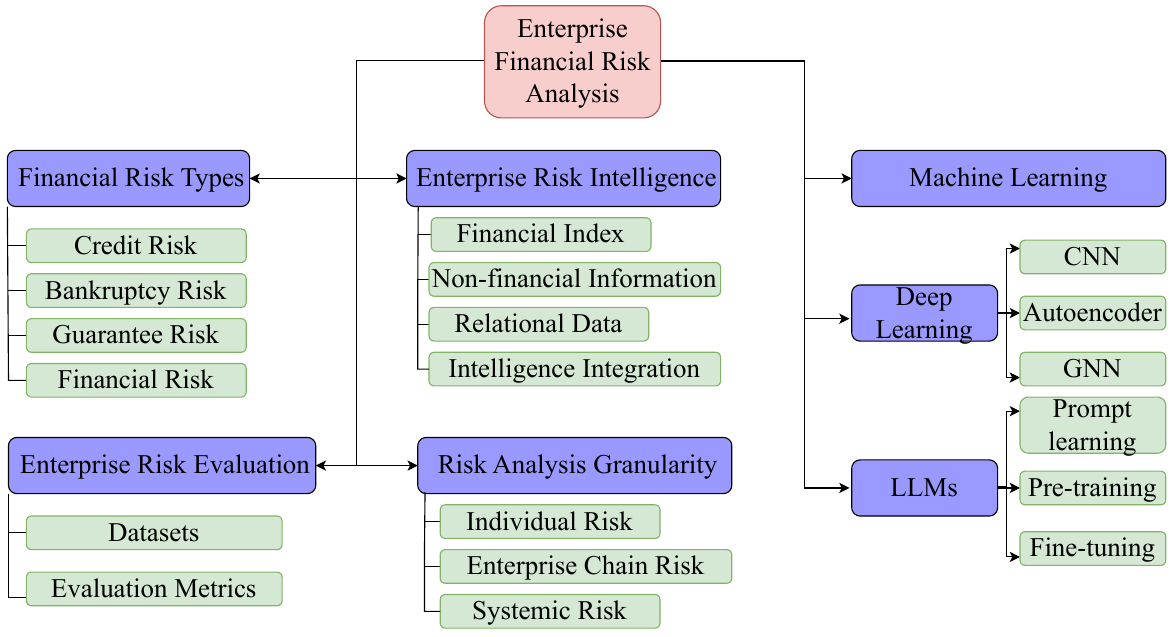}
    \caption{The framework of enterprise financial risk analysis.}
    \label{figure-framework}
\end{figure*}

Figure \ref{figure-publication-overview} and Figure \ref{figure-ration} show the number of relevant publications in the past thirteen years and their distribution in terms of different research directions. Due to the rapid expansion of the enterprise risk research based on the AI technologies, it is both necessary and challenging to comprehensively review the relevant studies. 
In this work, we systematically review a large body of relevant literature, aiming to comprehensively summarize the issues, methodologies, and key highlights in enterprise financial risk analysis, thereby helping readers gain a better understanding of the current research status and approaches. Figure \ref{figure-framework} presents the framework of this survey.

The contributions of this work are \underline{threefold}: (1)This survey provides a systematic and comprehensive review of existing research for enterprise financial risk. To the best of our knowledge, this is the first and only survey work on enterprise financial risk from the perspectives of Big Data and LLMs.
(2)We propose a novel and more comprehensive taxonomy for enterprise financial risk analysis. For each specific part, the contemporary
landscape is summarized and insightful analysis is provided.
 (3)We discuss the limitations of current research and propose five potential research
directions for enterprise financial risk analysis.

% There are many articles on enterprise risk, and a search of "enterprise risk" in google scholar gives about 56,000 results sorted by relevance. 
% It is very difficult for researchers who do not yet have knowledge of enterprise risk to quickly grasp the state of research today, so 
 
% We hope that this review will provide a basic overview for scholars to enter the field of enterprise risk. 
% Figure \ref{figure-EIGAT-Model} shows the distribution of our cited literature over time. 
% Since the time range of our references is too long, we only show the distribution of literature within the last 12 years here.
% \noindent \textbf{\textit{The framework:}}

The remainder of this article is organized as follows. 
Section \ref{section-Problem-Formalization} formalizes the problems of enterprise risk analysis from four different directions, including the risk types, analysis granularity, risk intelligence and evaluation metrics.
In Section \ref{section-methods}, we will systematically introduce the approaches used more frequently. 
Section \ref{section-spotlights} summarizes the spotlights of the most representative studies. 
% Section \ref{section-applications} introduces the application of enterprise risk analysis. 
Section \ref{section-futurework} provides guidance for future research directions. 
Finally, Section \ref{section-conclusion} provides a conclusion to the survey.
% we conclude the survey in Section \ref{section-conclusion}.

\section{Problem Formalization}
\label{section-Problem-Formalization}

\subsection{Financial Risk Types}
In this section, we review the previous research on enterprise risk analysis in terms of risk types, i.e., credit risk, bankruptcy risk, guarantee risk, and financial risk, which are summarized in Table \ref{table-enterprise-risk-types}. 
\subsubsection{Credit Risk} 
% loan default prediction

% \textbf{\textcolor{blue}{Wangmeiai}}
Enterprise credit risk refers to the potential for financial loss arising from the default of a counterparty or from changes in credit quality or ethical behavior during a transaction. The characteristics of enterprise credit risk are multifaceted and include comprehensiveness, bidirectionality, transitivity, diffusivity, cumulativeness, hiddenness, suddenness, and uncertainty. Credit risk is predominantly found in various credit transactions between financial institutions and enterprises, enterprises and other enterprises, as well as between governments and enterprises. Credit risk can originate from several sources, such as customer defaults or internal mismanagement. Moreover, it can also be triggered by external macroeconomic conditions or issues inherent to the financing firm itself. 
% Duffie \textit{et al.} \cite{duffie2007multi} demonstrate that the level and shape of a firm's default probability over time is influenced by its default distance. Belas \textit{et al.} \cite{belas2018impact} investigate the determinants of SME credit risk, with a particular focus on the influence of social and economic factors. Tang \textit{et al.} \cite{tang2010market} examine the interaction between market conditions and default risk, and how these dynamics affect firms' credit spreads. 

\subsubsection{Bankruptcy Risk}
Corporate bankruptcy risk refers to the likelihood that a company may be unable to fulfill its debt obligations or maintain sufficient funds to sustain normal operations, potentially leading to bankruptcy. There are many factors that can contribute to corporate bankruptcy, including the company's financial structure, governance framework \cite{Daily1994Bankruptcy}, corporate social responsibility practices, and macroeconomic conditions, among others.

\begin{table*}[t]
     \centering
    \caption{Literature comparison in terms of enterprise risk types.}
    \centering
    \label{table-enterprise-risk-types}
    \newcommand{\tabincell}[2]{\begin{tabular}{@{}#1@{}}#2\end{tabular}}
    \resizebox{0.98\textwidth}{!}{
    \begin{threeparttable}
      \begin{tabular}{lcccccccl}
        \toprule[1.1pt]
        \multirow{2}{*}{\textbf{\makecell{Enterprise Risk\\ Types}}} &  \multirow{2}{*}{\textbf{Literature}} &  \multicolumn{2}{c}{\textbf{Focus}} & \multicolumn{3}{c}{\textbf{Experiment}} &     \multicolumn{2}{c}{\textbf{Method}} \\
        \cmidrule(r){3-4}
        \cmidrule(r){5-7}
        \cmidrule(r){8-9}
        &  & \textbf{Industry}& \textbf{Country}&\textbf{Period} & \textbf{Size}  & \textbf{Metric} &  \textbf{Category} &  \textbf{Methodology}\\
        \midrule
        \multirow{2}{*}{\textbf{Credit Risk}}
        &\cite{Yin2020Evaluating} & Manufacturing & China  & 2015-2017 & 1,091& AUC, KS & SEM & MLR\\
        & \cite{zhu2016predicting}& All industries& China& 2012-2013&48 &AC, Type I, II error &ML &RSRAB
 \\
        \midrule
        \multirow{2}{*}{\textbf{Bankruptcy Risk}}& \cite{kou2021bankruptcy}&All industries &China & 2016-2018&3.5M &AUC & ML& TSMOFS \\
         & \cite{hosaka2019bankruptcy}	&All industries & Japan&2002-2016 &2,062 & AC& DL& CNN \\
        \midrule
         \multirow{2}{*}{\textbf{Guarantee Risk}}
         & \cite{cheng2020risk}	&All industries&Asia &2013-2016 & 0.11M& AUC & DL& DGANN\\
         & \cite{cheng2020contagious}&All industries& Asia&2013-2016 &0.11M &Macro-F1 score  & DL& DNN\\
        \midrule
        \multirow{2}{*}{\textbf{Financial Risk}}
         & \cite{zhai2021retracted}& High-tech industries&Worldwide  &2016-2020 &600 &AC & ML&PSO-BP\\
         & \cite{cao2022study} &E-commerce &China &- &3,617  &AC &DL &DL, DFT\\
     %   Guarantee Risk& \cite{}	& ... \\
        \bottomrule[1.1pt]
        \end{tabular}
		\begin{tablenotes}
			\footnotesize
			\item ML: The machine learning methods. SEM: Statistical econometric measurement. DL: The deep learning methods. AC: Average accuracy. AUC: Area under curve. CNN: Convolutional neural networks. DGANN: The dynamic graph attention neural network. DNN: The deep neural network. MLR: Multiple Linear regression. DFT: Data fusion technology. RSRAB: Random Subspace-Real AdaBoost. GA: The genetic algorithm. TSMOFS: Two-stage multiobjective feature-selection.
		\end{tablenotes}	
	\end{threeparttable}
        }
\end{table*}

\subsubsection{Guarantee Risk}
% \textbf{\textcolor{blue}{Zhu Yuansong}}
% Guarantee risk is the possibility that a credit guarantee agency will suffer losses due to various uncertainties in the course of its guarantee business operations. Depending on the financing needs of enterprises, guarantee methods are diversified, and common guarantee methods include equity swap guarantee\cite{wang2015entrepreneurial}, \cite{xiang2015investment}, related party guarantee\cite{berkman2009expropriation}, government guarantee\cite{wilcox2019government}, \cite{Sosin1980OnTV}, \cite{luo2016investment}, and mutual guarantee. The study of guarantee risk assessment and its system can help improve the guarantee capacity of SMEs and improve their financing difficulties.  \textbf{\textcolor{red}{This section will mainly introduce some theoretical achievements of guarantee risk.}}
Guarantee risk arises when a lender is unable to recover the principal and interest of a loan through either collateral or a guarantor due to the borrower's failure to repay on time or a default. At the firm level, Cowan \textit{et al.} \cite{cowan2015effect} demonstrate that partial credit guarantees significantly increase the default rates of insured loans, suggesting that guarantees influence firms' incentives to meet their repayment obligations. At the financial institution level, Wu \textit{et al.} \cite{wu2018Asset} develop a return model for bond securities, incorporating guarantee letters and proposing two key indicators to calculate the rate of return on securities.

\subsubsection{Financial Risk}
% Financial risk refers to the risk that an enterprise may not be able to fulfill its financial responsibilities and obligations as expected due to various unpredictable or uncontrollable factors in its financial activities, resulting in possible losses. \textcolor{red}{This section will review the existing literature from the aspects of financial risk warning, and influencing factors.} 

% Financial risk early warning involves risk identification, assessment, warning, and treatment using various methods based on enterprise financial data \cite{wei2022analysis, zhu2021financial,qiao2019enterprise}. These methods detect abnormal indicators, helping enterprises improve management and take appropriate actions.
% Financial risk is linked to board management, technological innovation investment, corporate social performance, and other factors. High-standard non-executive directors and cognitive conflict in decision-making reduce risk \cite{mcnulty2013boards}. Investment in innovation significantly lowers financial risk, especially for small and private enterprises \cite{su2022research}. Additionally, a negative correlation exists between corporate social performance and financial risk \cite{boubaker2020does,landi2022embedding}.

Financial risk refers to the potential for an enterprise to incur losses due to unpredictable or uncontrollable factors, leading to a discrepancy between actual and expected income in various financial activities. Financial risks can be categorized based on their origin, including liquidity risk, financing risk, and investment risk. The presence of such risks may result in fluctuations in the value of a firm's assets, reduced profitability, or a deterioration in financial health. The financial risks faced by enterprises are influenced by numerous factors, which can be broadly analyzed from two perspectives: internal and external \cite{wei2022analysis}. 
\subsection{Risk Analysis Granularity}
In this section, we review the previous research on risk analysis in terms of different granularity, i.e., individual enterprise risk, enterprise chain risk, and system risk, which are summarized in Table \ref{table-enterprise-risk-aspect}.

\begin{table*}[t]
    \centering
         \caption{Literature comparison in terms of risk analysis granularity.}
         \label{table-enterprise-risk-aspect}
         \newcommand{\tabincell}[2]{\begin{tabular}{@{}#1@{}}#2\end{tabular}}         
        \resizebox{\textwidth}{!}{           
           \begin{tabular}{lllcccccccl}
            \toprule[1.1pt]
            \multirow{2}{*}{\textbf{\makecell{Risk Analysis\\ Aspect}}} &  \multirow{2}{*}{\textbf{Literature}}  & \multicolumn{2}{c}{\textbf{Focus}} & \multicolumn{3}{c}{\textbf{Experiment}} & \multicolumn{2}{c}{\textbf{Methodology}} \\
            \cmidrule(r){3-4}\cmidrule(r){5-7}\cmidrule(r){8-9}
            & &  \textbf{Industry} &\textbf{Country} &\textbf{Period} & \textbf{Size}  & \textbf{Metric} &\textbf{Category} &\textbf{Methodology} \\
            \midrule
            \multirow{2}{*}{\textbf{\makecell{Individual\\ Enterprise}}} 
            & \cite{liang2015effect}&Banking&\makecell[c]{Australian, German\\China}&2000-2010&2,818&AC, Type I error&SEM&LDA, t-test, LR
\\
            & \cite{ambulkar2015firm} &Manufacturing&India&2011-2012&205&CFA&SEM&Structural equation\\
           \midrule
           \multirow{2}{*}{\textbf{\makecell{Enterprise \\Chain}}}          
            %& \cite{oh2018time}&Financial&\makecell[c]{North America,\\ Europe}&2004-2010&121&P-value&SEM&Dynamic copula model\\
            & \cite{du2021dynamic}&All industries&French&1997-2015&10,000&AC, F-measure&ML&Self-organizing neural network\\
            &\cite{son2019data}&All industries&Poland&2008-2016&10,000&AC, AUC, Recall&ML&XGBoost, NN\\      
            \midrule
           \multirow{2}{*}{\textbf{Systemic}}
           % &\cite{shahzad2022pandemic}&Tourism&US&2018-2020&95&TRSI, SII&SEM&TALIS\\
            & \cite{zhang2020government}&Commerce&China&2007-2016&86&Z-score&SEM&Difference estimation method\\
            % & \cite{kritzman2011principal}&Financial markets&US&1998-2010&116&P-value&SEM&PCA\\
            & \cite{ramiah2013does}&All industries&European, US&1999-2018&1,770&CARs, CAARs, ARV&SEM&CAPM\\     
          \bottomrule[1.1pt]
           \end{tabular}
         }
    % \end{center}
\end{table*}

%%%%%
\subsubsection{Individual Enterprise Risk}
% \textbf{\textcolor{blue}{Yuanyiying}}
Individual enterprises encounter numerous risks in their operations and are subject to various pressures in the course of management activities. In the context of trade, multiple enterprises often collaborate to maximize overall benefits. In this section, we analyze enterprise-specific risks by industry type, noting that current research predominantly emphasizes the industrial and manufacturing sectors.

In the industrial sector, In the industrial sector, some studies focus on comparative analyses of traditional statistical methods for financial distress classification and prediction, including Linear Discriminant Analysis (LDA) and Neural Networks (NN). In the manufacturing and construction industries, novel methods have been proposed to enhance the predictability of corporate bankruptcy and insolvency risks, including Self-Organizing Maps (SOM) and Z-score models. For the financial industry, previous research has primarily focused on banking, highlighting the importance of analyzing the risk propagation process between banks.
%Among them, it is extremely important to analyze the risk contagion of banks and study the risk propagation process \cite{huang2013cascading, Acemoglu2015Systemic}. 

%%%%%
\subsubsection{Enterprise Chain Risk}
% \textbf{\textcolor{blue}{Zhu Yuansong}}
% In a large financial social system, companies tend to collaborate and influence each other. We can call it an enterprise chain if there is a linkage and some kind of transmission relationship between multiple enterprises. Companies at different points play a role in the formation and stability of the enterprise chain (supply, sales, etc.). Financing, investment, production and operation activities are always happening in the enterprise chain, which are accompanied by profits and risks. \textbf{\textcolor{red}{This section provides an introduction to the main types of enterprise chain risks that may be encountered in the above-mentioned financial activities and lists the contributions of some scholars in the field, mainly including supply chain, guarantee chain, transaction chain and credit chain.}}

%In complex financial ecosystems, companies often collaborate and exert mutual influence. When multiple enterprises are linked by transmission (association) relationships, we can refer to this as an enterprise chain. This section introduces the primary types of risks that may arise within enterprise chains during financial activities.
Enterprise chain is a macro concept that essentially refers to a network of companies linked by intrinsic connections, encompassing both structural and value dimensions. Enterprise chains can be classified into various types, including supply chains, guarantee chains, transaction chains, and credit chains. This section provides a risk analysis for different types of enterprise chains.

In supply chain research, Heckmann \textit{et al.} \cite{heckmann2015critical} identify key characteristics that define, quantify, and model supply chain risks through an extensive literature review. In the context of guarantee chains, Cheng \textit{et al.} 
% \cite{cheng2020delinquent} explore the prediction of successive default events in network-guaranteed loans. 
Regarding transaction chains, Kou \textit{et al.} \cite{kou2021bankruptcy} introduce a bankruptcy prediction model for SMEs, using transaction data and payment network-based variables instead of traditional financial accounting data.
% The following three papers analyze and study supply chain risks from different perspectives. Heckmann \textit{et al.} \cite{heckmann2015critical} based on a literature review, identify core characteristics used to define, quantify and model supply chain risks. Babich \textit{et al.} \cite{babich2007competition} finds that competition among suppliers in a supply chain affects the equilibrium wholesale price. Blome \textit{et al.} \cite{blome2011supply} presents a set of propositions on how firms manage supply risk in a financial crisis. In the context of guarantee chains, Cheng \textit{et al.}\cite{cheng2020delinquent}investigates the prediction of successive default events for network-guaranteed loans. Regarding the study of transaction chains, without the need for financial (accounting) data, Kou \textit{et al.} \cite{kou2021bankruptcy} proposes a model for predicting SME bankruptcy using transaction data and payment network-based variables. As for credit chains, Battiston \textit{et al.}\cite{Battiston2007CreditCA} propose a simple production network model by linking supplier-customer relationships involving trade credit extensions.

% \noindent \textbf{\textit{Supply Chain Risk.}} 

% \noindent \textbf{\textit{Guarantee Chain Risk.}} \cite{cheng2020contagious}  

% Credit risk evaluation for loan guarantee chain in China \cite{Meng2015Credit}

%%%%%
\subsubsection{Systemic Risk}
% \textbf{\textcolor{blue}{SunXiaoying}}
Systemic risk refers to the risk that affects an entire market or market segment and is caused by factors beyond the control of any specific company or individual. Below, we provide an overview of systemic risk based on its sources. Recent studies indicate that systemic events—such as COVID-19 and terrorist attacks—can have profound impacts on the global economy. The formulation and implementation of national policies also play a critical role in shaping economic trajectories \cite{zhang2020government}. Moreover, disturbances within financial networks may trigger cascading effects, potentially leading to systemic collapses.

%%%%%%%%%%%
\subsection{Enterprise Risk Intelligence}
\begin{figure*}[t]
    \centering
    \includegraphics[width=0.8\textwidth]{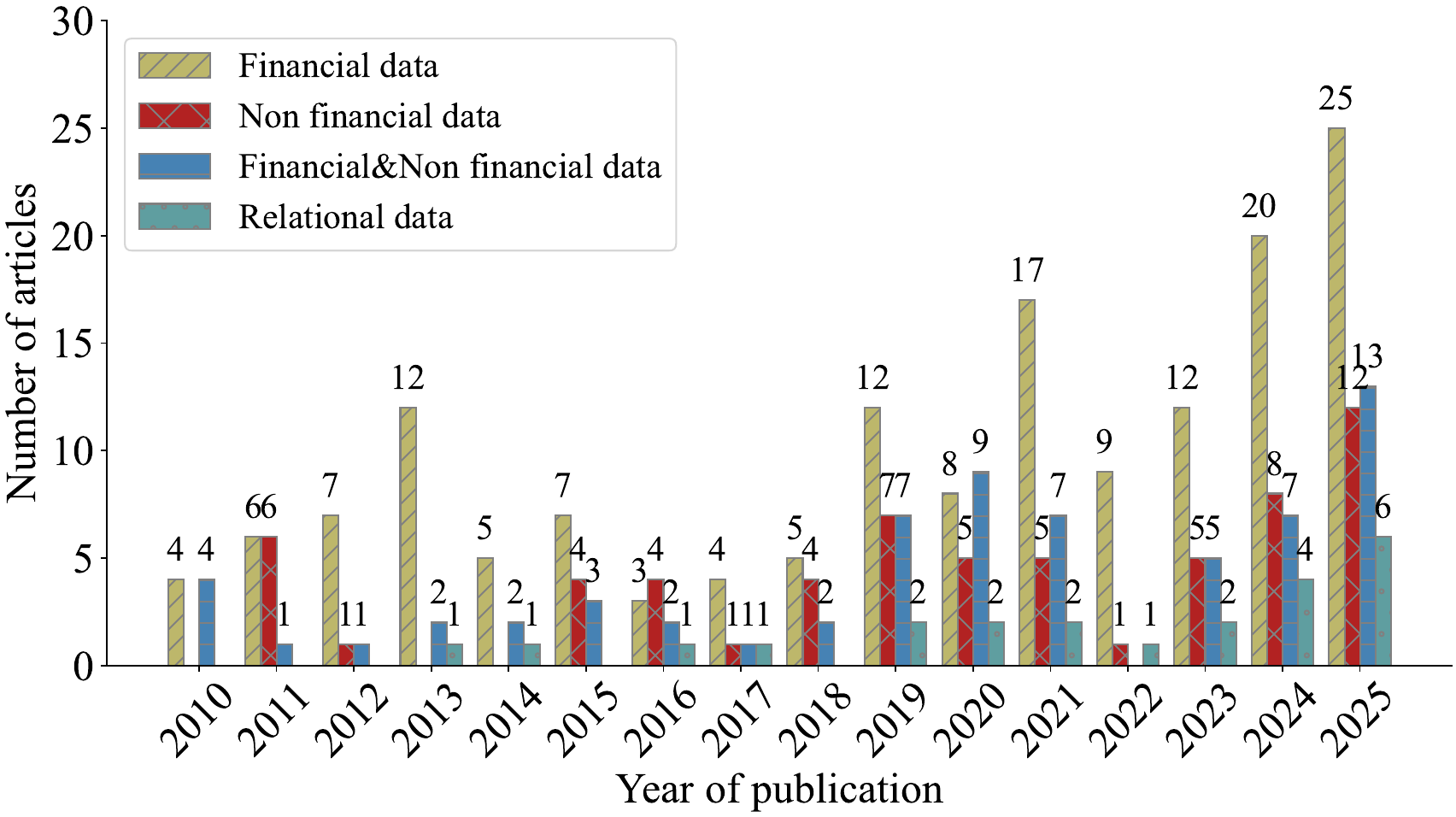}
    \caption{Publication trends of enterprise risk intelligence over the years.}
    \label{figure-publication-risk-intelligence}
\end{figure*}

Table \ref{table-enterprise-risk-intelligence} presents previous studies on risk analysis in terms of enterprise risk intelligence, i.e., financial index, non-financial textual information, relational data, and intelligence integration. And Figure \ref{figure-publication-risk-intelligence} shows the publication treads of enterprise risk intelligence over the years.

\begin{table*}[t]
    \centering
         \caption{Literature comparison in terms of enterprise risk intelligence.}
         \label{table-enterprise-risk-intelligence}
         \newcommand{\tabincell}[2]{\begin{tabular}{@{}#1@{}}#2\end{tabular}}
        \resizebox{0.95\textwidth}{!}{
           \begin{tabular}{llclccccl}
            \toprule[1.1pt]
            \multirow{2}{*}{\textbf{Enterprise Risk Intelligence}} &  \multirow{2}{*}{\textbf{Literature}} & \multicolumn{3}{c}{\textbf{Experiment}} & \multicolumn{2}{c}{\textbf{Methodology}} \\
            \cmidrule(r){3-5}\cmidrule(r){6-7}
            & & \textbf{Period} & \textbf{Size}  & \textbf{Metric} & \textbf{Category} & \textbf{Methodology}  \\
            \midrule
            \multirow{2}{*}{\textbf{Financial Index}}
            & \cite{cheng2022regulating}&2004-2015&0.83 M&AC&DL&GNN\\
            & \cite{ribeiro2019shaping}&2007&250&AUC&DL&Graph Pattern Mining\\
           \midrule
           
           \multirow{2}{*}{\textbf{\makecell{Non-financial Textual\\ Information}}}
            & \cite{bai2020innovate}&1980-2009&16,966&P-value&SEM&Discrete hazard model\\
            % & \cite{niu2020iconviz}&-&20000&-&ML&LR, SVM\\
            & \cite{hu2020loan}&-&1.5 M&KS, AUC&DL&AMG-DP model\\
            % & \cite{ye2020financial}&2015-2018&6,494&MAE&DL&MR-QA\\
            \midrule
           \multirow{3}{*}{\textbf{Relational Data}}
            & \cite{bi2023Company}&2019-2020&4,040&AUC&DL&TH-GNN \\
            % & \cite{wang2021ignorance}&2011-2018&0.43 M&-&SEM&Three-factor model\\
            & \cite{Zhao2022Bankruptcy}&2000-2021&0.11 M&AC, Recall, F1-score&DL&GNN\\
            & \cite{yang2021Financial}&-&1 M&AUC&DL&ST-GNN \\
            \midrule  
           \multirow{2}{*}{\textbf{Intelligence Integration}}
            % & \cite{alam2021corporate}&2001-2018&0.64 M&AC&ML&BPN\\
            & \cite{Wang2021MultiviewGL}&2011-2019&924&AC&DL&M-GL\\
            & \cite{duan2021bank} &2020&1,584&P-value&SEM&DCC-GARCH model\\
            \bottomrule[1.1pt]
           \end{tabular}
         }
    % \end{center}
\end{table*}

%%%%%
\subsubsection{Financial Index}
Financial indicators are relative metrics that enterprises use to summarize and evaluate their financial conditions and operational performance, including indicators of solvency, operating capacity, and profitability. Common solvency indicators include the asset–liability ratio, quick ratio, and working capital. These indicators serve as important tools for studying the factors influencing bankruptcy and analyzing risk characteristics. Profitability indicators can be used to measure a company's profit level and assess the stability of supply and demand, thereby helping banks predict corporate creditworthiness, reduce financial distress, and alleviate liquidity shortages in the interbank market.

% The first work \cite{Altman1968Financial}

% Node attributes: registered capital, number of employees, financial status; Edge attributes: guarantee amount, loan amount, loan interest. \cite{Cheng2022Regulating}

% financial ratios, economic indicators, and technology evaluation factors \cite{Kim2010SupportVM}

% Profitability, Productivity, and Liquidity \cite{lee2013multi}

% Cash flow \cite{gombola1987cash,jones2016cash,zhang2020structured}

%%%%%%
\subsubsection{Non-financial Information}

% \textbf{\textcolor{blue}{SunXiaoying}}
\noindent 

\textbf{\textit{Non-financial variables:}} Non-financial variables play a crucial role in evaluating a company’s success across multiple dimensions, including operations, customer relationships, and employee engagement. Key non-financial indicators related to customers include metrics such as customer churn rates and personal ratings. In terms of employee-related non-financial factors, board structure has become an important area of research. In addition, non-financial variables related to business operations typically encompass business cycles and macroeconomic indicators \cite{du2021dynamic}.

%%%
\noindent 

\textbf{\textit{Textual Information:}}
Existing research leverages news reports and sentiment analysis to assess risk by analyzing textual information in news articles and qualitative risk disclosures in corporate filings. In addition, legal documents, earnings call transcripts, and interview transcripts also provide valuable insights for risk assessment. For example, Yin \textit{et al.} \cite{yin2020evaluating} integrate data on legal actions taken by creditors to recover outstanding debts, company filing history, consolidated audit opinions, firm-specific characteristics, and legal judgments to develop models predicting corporate default risk.

%%%%%
\subsubsection{Relational Data}
% \zhaolanqi
Relational data is used to characterize the connections between entities, including equity relationships, cooperative relationships, supply chain relationships, management relationships, and so on. Researchers utilize relational data to study the network of relationships among firms. Moreover, relational data is particularly prevalent in the study of supply chain networks. By collecting supply chain relationship data between enterprises—including direct suppliers, sub-tier suppliers, and interactions among suppliers—researchers can construct multi-tiered supply chain networks.

%%%%%%%%%%%
\subsubsection{Intelligence Integration}

% \edit{liangqiaoyue}\\
In the assessment of corporate risk, different kinds of information are captured. Generally, financial data is integrated with non-financial data to obtain comprehensive information. Non-financial information, like firm size, corporate governance indexes, and audit opinions, has already been demonstrated to be practical in risk prediction \cite{shen2020dynamic}. Due to varied research goals, the specific types of non-financial information utilized in risk assessment may vary. 
% Furthermore, there is a growing trend in risk management towards uniting relational data with financial data. For example, Cleofas \textit{et al.}  \cite{cleofas2016financial} combines accounting statements with business information, bank customer data, and credit applicant records to predict bankruptcy. Such data depicts the relationship between a company and linked enterprises \cite{gofman2022trade}, or the relationship between a company and banks \cite{chan2013does}. Integration also happens among the three types of data mentioned above. For instance, Zhao \textit{et al.} \cite{zhao2022stock} utilize transaction data, web news as well as company relations and executives’ information to predict the stock movement. 
%\textcolor{red}{In summary, we anticipate that in future research, there will be an increasing exploration of data integration, leading to significant improvements in the study of enterprise risks.}

%%%%%%%%%%%
\subsection{Enterprise Risk Evaluation}
%%%%%
\subsubsection{Datasets}

We present the various datasets used  for enterprise risk analysis in Table \ref{Features-statistics}. These datasets encompass different types and domains of information, providing scholars with rich materials to analyze and evaluate enterprises' financial risks. The most commonly utilized datasets in previous research include SME datasets, financial datasets, stock datasets, rating datasets, and credit guarantee datasets. Traditionally, most statistical econometric models have relied heavily on financial and text data to assess enterprise risk. However, with advancements in machine learning and deep learning approaches, a broader range of datasets is now employed. For instance, bank loan and credit guarantee data offer valuable insights into financing and credit risks.
% \zhaolanqi
\begin{table*}[t] % [!ht]表格在文本中放置的位置参数（努力放在当前位置，实在放不下，将放在下一页的顶部）

\centering % 表格整体居中
    \caption{Description of datasets.}
    \label{Features-statistics}
    \newcommand{\tabincell}[2]{\begin{tabular}{@{}#1@{}}#2\end{tabular}}
    \resizebox{0.99\textwidth}{!}{
    \begin{tabular}{c|c|c|c|c} 
    \toprule[1.1pt]
     \textbf{Datasets}&\textbf{Tasks} & \textbf{Descriptions} & \textbf{Period} & \textbf{Size}\\
    \midrule
        FinDABench \cite{liu2024fidabench}&\tabincell{c}{Analytical reasoning ability}&\tabincell{c}{perform financial indicator calculation \\and corporate sentiment risk assessment.}&-&\tabincell{c}{15,200 training instances\\ 8,900 test instances}\\
    \midrule
    SMEsD \cite{Zhao2022Bankruptcy} &Bankruptcy Risk Prediction  &\tabincell{c}{This dataset is a benchmark dataset constructed by \\collecting SME real data from multiple sources.}&2014-2020&4,229 \\ 
    \midrule 
CFLUE \cite{zhu2024CFLUE}&\tabincell{c}{knowledge assessment and \\application assessment}&\tabincell{c}{
With over 38K multiple-choice questions and \\16K instances across generation tasks}&-&38k, 16k \\ 
 \midrule 
   REFinD \cite{Kaur2023REFinD}&Relation Extraction&\tabincell{c}{29K instances and 22 relations amongst 8 types of \\entity pairs over financial documents}&2016-2017&28,676\\ 
   \midrule
BIGPATENT \cite{sharma2019bigpatent}&Extractive summarization&\tabincell{c}{Contains 1.3 million U.S. patent document records along with \\human-written abstractive summaries.
}&1971-&1.3 million \\ 
   \midrule  
   \tabincell{c}{FINQA \cite{chen2021finqa}} &\tabincell{c}{Question Answering}&\tabincell{c}{Based on the publicly available earnings reports of S\&P 500\\ companies from 1999 to 2019 collected in the FinTabNet dataset.}&1999-2019&828,1\\ 
  \bottomrule[1.1pt]
\end{tabular}
}
\end{table*}
%%%%%
\subsubsection{Evaluation Metrics}
% \zhaolanqi
% \todo{Add a table}
% \edit{Add a paragraph as an introduction to this section}
% \par The evaluation index and evaluation system of the model is an important link in the modeling process. For different types of projects and models, different evaluation index and system should be reasonably selected. In this section, we selected four evaluation indicators for a detailed introduction.
%Evaluation metrics are crucial in the modeling process. Appropriate indices should be chosen for different tasks and models. In this section, three evaluation indicators are discussed in detail.
Evaluation metrics are crucial in the modeling process, as appropriate indices must be selected based on the specific tasks and models involved. In this section, we discuss three evaluation indicators in detail.
% and Table \ref{Statistic Main Formular} presents the main formulae for calculating these indicators.

\noindent

\textbf{\textit{Kolmogorov-Smirnov statistic (KS):}} 
% The K-S test is a goodness-of-fit test for exploring the distribution of continuous random variables, using sample data to infer whether the overall population from which the sample is drawn follows a certain theoretical distribution. The KS test can be used to assess the discriminatory performance of a predictive model. Chang \textit{et al.} \cite{Yin2020Evaluating} performed 10 independent 10-fold cross-validations against KS in order to predict the recognition performance of the model and obtained 100 performance estimates.
%The K-S test examines the distribution of continuous random variables and evaluates model's discrimination performance. Using KS, Chang \textit{et al.} \cite{Yin2020Evaluating} conduct 10 independent 10-fold cross-validations, predicting model recognition performance and obtaining 100 performance estimates.
The K-S test examines the distribution of continuous random variables and evaluates a model's discrimination performance. Using the K-S test, Chang \textit{et al.} \cite{Yin2020Evaluating} conduct 10 independent 10-fold cross-validations to assess model recognition performance.
%resulting in 100 performance estimates.
\noindent

\textbf{\textit{AUC:}} 
%The AUC curve, under the ROC curve and enclosed by axes, measures separability and learner quality. It is an important measure for financial risk prediction \cite{peng2011empirical}, where a larger AUC indicates a better classifier. AUC compares prediction effectiveness for hard, soft, and integrated features with various approaches. For default risk prediction, which is an unbalanced classification problem, AUC reflects the classifier's learning effect more fully \cite{Huang2021TwostageAI,sun2018imbalanced}.
The AUC measures the separability and quality of a learner by assessing the ROC curve, which is enclosed by the axes. It is a crucial metric for financial risk prediction, where a larger AUC indicates a better classifier. The AUC facilitates the comparison of prediction effectiveness for hard, soft, and integrated features across various approaches. In the context of default risk prediction, which presents an unbalanced classification challenge, the AUC more comprehensively reflects the classifier's learning effectiveness.

\noindent 

\textbf{\textit{Kappa measure: }} 
% The Kappa coefficient is an index to measure classification accuracy based on confusion matrix calculation. It takes into account the possibility of the raters agreeing by chance, thus being more reflective of the classifier’s performance than simple accuracy or precision/recall. It is a good evaluation index to measure the classification effect of data with unbalanced samples. The higher the Kappa coefficient is, the better the prediction effect will be 
%Using confusion matrices, the Kappa coefficient measures classification accuracy, considers chance agreement, and offers a better performance reflection than accuracy or precision and recall. It is a decent option for assessing unbalanced data classification, with higher coefficients indicating better prediction \cite{shen2020dynamic}.
Using confusion matrices, the Kappa coefficient measures classification accuracy while accounting for chance agreement. It provides a more reliable reflection of performance compared to accuracy, precision, and recall. The Kappa coefficient is particularly useful for assessing unbalanced data classification, with higher values indicating better predictive performance \cite{shen2020dynamic}.
%%%%%%%%%%%%%%%%%%%%%%%%%%%%%%%%%%%%%%%%%%%%%%%%%%%%%%%%%
\section{Methodologies}
\label{section-methods}
%Table \ref{table-analysis-models} presents the previous studies on risk analysis in terms of analysis models, i.e., statistical model, machine learning model, deep learning model and hybrid model. Figure \ref{timeline} exhibits a chronological overview of influential enterprise risk analysis method. 
Existing studies primarily employ machine learning models, deep learning models, and LLMs for enterprise financial risk analysis. Figure \ref{timeline} provides a chronological overview of significant methods used in enterprise risk analysis. 
% Additionally, Figure \ref{figure-publication-analysis-model} illustrates the publication trends related to these analysis models over the years.
%In Figure \ref{timeline}, we plot the distribution of representative works of some major models on the timeline. 
%And Figure \ref{figure-publication-analysis-model} shows the publication trends of analysis models over the years.

% 画时间轴的部分
\begin{figure*}[htb]
\tikzstyle{descript} = [text = black,align=center, minimum height=1.8cm, align=center, outer sep=0pt,font = \scriptsize] %这一行不懂啥意思
\tikzstyle{activity} =[align=center,outer sep=1pt]
\resizebox{1\textwidth}{!}{
\begin{tikzpicture}[very thick, black]
\small
%% Coordinates
\coordinate (O) at (0,0); % Origin
% \coordinate (P1) at (4,0);
% \coordinate (P2) at (8,0);
% \coordinate (P3) at (12,0);
\coordinate (F) at (13,0); %End

% \coordinate (1990) at (0,0); 
% \coordinate (2000) at (2,0); 
% \coordinate (2005) at (4,0); 
% \coordinate (2010) at (0,0); 
% \coordinate (2015) at (0,0); 
\coordinate (2017) at (0,0); 
\coordinate (2019) at (2,0); 
\coordinate (2021) at (4,0); 
\coordinate (2022) at (6,0); 
\coordinate (2023) at (8,0);
\coordinate (2024) at (10,0);
\coordinate (2025) at (12,0);
% \coordinate (2027) at (14,0);

%画文献箭头
%深度学习
\draw[-,very thick,color=blue] ($(2019)+(0.6,0)$) -- ($(2019)+(0.6,0.6)$) node [above=0pt,align=center,blue] {\cite{hosaka2019bankruptcy}\\CNN}node[circle,fill,inner sep=1.15pt] at ($(2019)+(0.6,0.6)$) {};%第1根蓝线
\draw[-,very thick,color=blue] ($(2019)+(1.5,0)$) -- ($(2019)+(1.5,0.5)$) node [above=0pt,align=center,blue] {\cite{ye2020Financial}\\MRQ\\\&AAN}node[circle,fill,inner sep=1.15pt] at ($(2019)+(1.5,0.5)$) {};%第2根蓝线
% \draw[-,very thick,color=blue] ($(2019)+(1.7,0)$) --  ($(2019)+(1.7,-0.5)$) node [below=-8pt,align=center,blue] {\cite{smiti2020bankruptcy}\kern+0.9cm\\BSM-SAES}node[circle,fill,inner sep=1.15pt] at ($(2019)+(1.7,-0.5)$) {};%第3根蓝线
\draw[-,very thick,color=blue] ($(2019)+(1.1,0)$) -- ($(2019)+(1.1,-1.1)$) node [below=0pt,align=center,blue] {\cite{Soui2020Bankruptcy}SAE}node[circle,fill,inner sep=1.15pt] at ($(2019)+(1.1,-1.1)$) {};%第4根蓝线
\draw[-,very thick,color=blue] ($(2021)+(0.9,0)$) -- ($(2021)+(0.9,-0.6)$) node [below=-1pt,align=center,blue] {\cite{yang2021Financial}\\ST-GNN}node[circle,fill,inner sep=1.15pt] at ($(2021)+(0.9,-0.6)$) {};%第5根蓝线
\draw[-,very thick,color=blue] ($(2022)+(0.3,0)$) -- ($(2022)+(0.3,-0.5)$) node [below=-1pt,align=center,blue] {\cite{cheng2022regulating}\\iConReg}node[circle,fill,inner sep=1.15pt] at ($(2022)+(0.3,-0.5)$) {};%第6根蓝线

% \draw[-,very thick,color=blue] ($(2021)+(0.1,0)$) -- ($(2021)+(0.1,2.5)$) node [above=-1pt,align=center,blue] {\cite{yang2021financial}\\STAG-NN};%第5根蓝线
\draw[-,very thick,color=blue] ($(2022)+(0.7,0)$)-- ($(2022)+(0.7,0.8)$) node [above=-2pt,align=center,blue] {\cite{bi2023Company}\\TH-GNN}node[circle,fill,inner sep=1.15pt] at ($(2022)+(0.7,0.8)$) {};%第7根蓝线
\draw[-,very thick,color=blue] ($(2023)+(0.15,0)$) -- ($(2023)+(0.15,0.3)$) node [above=1.5pt,align=center,blue] {\cite{bi2023Predicting}\\KT-GNN}node[circle,fill,inner sep=1.15pt] at ($(2023)+(0.15,0.3)$) {};%第8根蓝线
% \kern-0.5cm
% %传统模型\kern-0.5cm
\draw[-,very thick,color=red] ($(2023)+(0.4,0)$) -- ($(2023)+(0.4,-0.45)$) node [below=0pt,align=center,red] {\cite{xie2023PIXIU}\\PIXIU};%第1根红线
\draw[-,very thick,color=red] ($(2023)+(1,0)$) -- ($(2023)+(1,1)$) node [above=0pt,align=center,red] {\cite{wu2023Bloomberggpt}\\BloombergGPT};%第1根红线
\draw[-,very thick,color=red] ($(2023)+(1.6,0)$) -- ($(2023)+(1.6,-0.4)$) node [below=0pt,align=center,red] {\cite{yang2023fingpt}\\FinGPT};%第1根红线
\draw[-,very thick,color=red] ($(2024)+(0.9,0)$) -- ($(2024)+(0.9,0.3)$) node [above=0pt,align=center,red] {\cite{zhang2024when}\\StockAgent};%第1根红线
\draw[-,very thick,color=red] ($(2024)+(1.5,0)$) -- ($(2024)+(1.5,-0.2)$) node [below=0pt,align=center,red] {\cite{zhang2024multimodal}\\FinAgent};%第1根红线
\draw[-,very thick,color=red] ($(2025)+(0.6,0)$) -- ($(2025)+(0.6,0.6)$) node [above=0pt,align=center,red] {\cite{li2025Investorbench}\\INVESTORBENCH};%第1根红线
\draw[-,very thick,color=ForestGreen] ($(2017)+(0.5,0)$) -- ($(2017)+(0.5,-0.65)$) node [below=-2pt,align=center,ForestGreen] {\cite{zorn2017cure}\\LR}node[circle,fill,inner sep=1.15pt] at ($(2017)+(0.5,-0.65)$) {};%第5根绿线
\draw[-,very thick,color=ForestGreen] ($(2017)+(0.8,0)$) -- ($(2017)+(0.8,0.8)$) node [above=0pt,align=center,ForestGreen] {\cite{ai2017optimal}\\DT}node[circle,fill,inner sep=1.15pt] at ($(2017)+(0.8,0.8)$) {};%倒数第3根绿线
\draw[-,very thick,color=ForestGreen] ($(2019)+(1.7,0)$) -- ($(2019)+(1.7,-0.6)$) node [below=0pt,align=center,ForestGreen] {\cite{uddin2022leveraging}RF}node[circle,fill,inner sep=1.15pt] at ($(2019)+(1.7,-0.6)$) {};%倒数第2根绿线
\draw[-,very thick,color=ForestGreen] ($(2021)+(0.7,0)$) -- ($(2021)+(0.7,0.5)$) node [above=0pt,align=center,ForestGreen,xshift=-0.1cm] {\cite{du2021Dynamic}NN}node[circle,fill,inner sep=1.15pt] at ($(2021)+(0.7,0.5)$) {};%倒数第1根绿线
\draw[->] (O) -- (F);
% Ticks
\foreach \x in {0,...,12}   %坐标轴标记点
\draw(\x cm,3pt) -- (\x cm,-1pt);   % 坐标轴标记点上下长度
%% Labels
\foreach \i \j in {0/2017,2/2019,4/2021,6/2022,8/2023,10/2024, 12/2025}{
	\draw (\i,0) node[below=5pt] {\j} ;
}
\end{tikzpicture}
}
\caption{Chronological overview of methods to studying enterprise risk. Methods in \textcolor{ForestGreen}{green}, \textcolor{blue}{blue} and \textcolor{red}{red} are machine learning methods, deep learning methods and large language model respectively.}
\label{timeline}
\end{figure*}

\subsection{Machine Learning Methods}

% \begin{figure*}[t]
%     \centering
%     \includegraphics[width=0.95\textwidth]{figures/chapter3.pic/ml4_3.png}
%     \caption{\textbf{The general pipeline of machine learning models.}}
%     \label{figure-ML}
% \end{figure*}

Machine learning methods are highly popular in enterprise risk analysis. Compared to traditional statistical approaches, machine learning offers several advantages, including greater adaptability, enabling it to effectively handle nonlinear relationships and high-dimensional data. Additionally, machine learning models demonstrate excellent scalability and model interpretability. For instance, Iturriaga \textit{et al.}. \cite{iturriaga2015bankruptcy} combined multilayer perceptrons with SOM in the financial domain. The performance of neural networks can be further enhanced through improved training methods, optimized network architectures, or the use of higher-quality input data.

\subsection{Deep Learning Methods}

% \begin{figure*}[t]
%     \centering
%     \includegraphics[width=0.95\textwidth]{figures/dl_plus.png}
%     \caption{\textbf{The general pipeline of deep learning models.}}
%     \label{figure-DL}
% \end{figure*}
%by liang: 
Deep learning methods overcome the limitations of manually designed features, enabling more effective financial risk analysis. This section discusses the various applications of three types of deep learning models—Convolutional Neural Networks, Autoencoders, and Graph Neural Networks—in enterprise risk analysis.

\subsubsection{Convolutional Neural Networks}
CNNs are advanced deep learning models that leverage unique features such as local connectivity and weight sharing. These unique features make CNNs highly effective in extracting significant features from complex financial data, revealing obscured patterns and trends that can facilitate precise predictions and informed decision-making in the market. Leveraging this capability, Wei \textit{et al.} \cite{wei2022analysis} extract both structured data and unstructured text from corporate annual reports to develop a financial risk prediction system utilizing CNN and LSTM. Similarly, Hosaka \textit{et al.} \cite{hosaka2019bankruptcy} transform financial scale data into grayscale images, subsequently applying CNNs for bankruptcy prediction.

\subsubsection{Autoencoder}
Autoencoder is an unsupervised neural network model that aims to compress input data into a lower-dimensional feature space, subsequently reconstructing the original data from this representation. This makes autoencoder valuable for enterprise financial risk analysis \cite{soui2020bankruptcy,smiti2020bankruptcy}. For instance, Soui \textit{et al.} \cite{soui2020bankruptcy}  employ Stacked Autoencoders (SAE) to identify the most important features.
% Smiti \textit{et al.} \cite{smiti2020bankruptcy} introduce a novel approach called BSM-SAES, which integrates a boundary synthesis minority oversampling technique with a stacked autoencoder based on the softmax classifier. 

% \subsubsection{Attention neural network}
% Attention neural network is a model that introduces an attentional mechanism into a neural network, which allows the neural network to focus on more important information by assigning different weights to different parts of the input. Ye et al. \cite{ye2020financial} used the attention mechanism to construct a Multi-Round Q\&A Attention Network model and used it to analyze earnings conference call transcripts.

\subsubsection{Graph Neural Networks}
%GNN is a deep learning method based on graph structure that transforms nodes and edges into a dense vector form and employs the topological information of the graph for computation and inference. And GNN is commonly used for node classification in enterprise risk analysis. For instance, Yang \textit{et al.} \cite{yang2021financial} put forward an innovative framework for financial risk analysis based on supply chain graph mining. Meanwhile, Bi \textit{et al.} \cite{bi2022company} propose TH-GNN, which is applied to tribe-style graphs, with the first level encoding the structural patterns of tribes through contrastive learning and the second level diffusing information based on inter-tribal relationships. Then Bi \textit{et al.} \cite{bi2023predicting} present KT-GNN, which simulates the transfer of knowledge from vocal nodes to silent nodes by transferring the distribution during message passing and representation learning. It is also used in the financial risk assessment scenario. Zhang \textit{et al.} \cite{zhang2024collaborative} proposed a novel Heterogeneous Graph Co-Attention Network for corporate default risk assessment. By utilizing collaborative metapaths, the model extracts risk features based on a co-attentive aggregation mechanism.
GNNs are deep learning methods that utilize graph structures by converting nodes and edges into dense vector representations, leveraging the topological information of the graph for computation and inference. GNNs are frequently applied in enterprise risk analysis, particularly for node classification \cite{yang2021Financial,bi2023Company,zhang2024collaborative,yuan2025dynamic}. For instance, Yang \textit{et al.} \cite{yang2021Financial} propose a framework for financial risk analysis that leverages supply chain graph mining. In a related effort, Bi \textit{et al.} \cite{bi2023Company} introduce TH-GNN, designed for tribe-style graphs, where the first level encodes the structural patterns of tribes using contrastive learning, and the second level disseminates information based on inter-tribal relationships. 
% Additionally, Zhang \textit{et al.} \cite{zhang2024collaborative} propose a novel Heterogeneous Graph Co-Attention Network for corporate default risk assessment, which utilizes collaborative metapaths to extract risk features through a co-attentive aggregation mechanism. 
DGNN-SR \cite{yuan2025dynamic} introduces a multi-view time encoder and an adaptive re-weighting strategy, enabling the simultaneous encoding of dynamic transaction relationships and static fund transfer relationships for credit risk assessment.

\subsection{Large Language Models}
\begin{figure*}[t]
    \centering
    \includegraphics[width=0.92\textwidth]{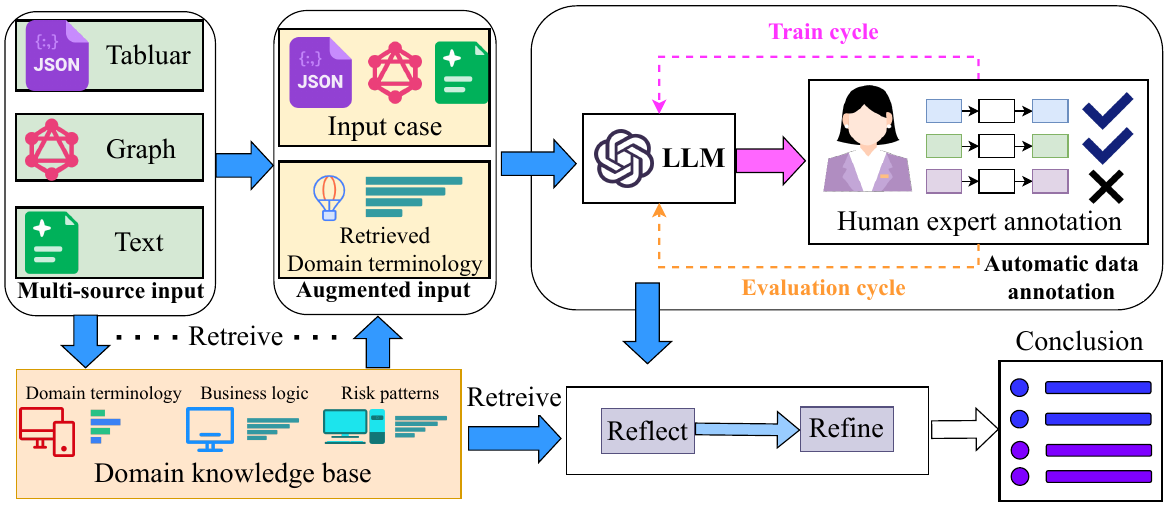}
    \caption{The workflow of LLMs in enterprise risk analysis.}
    \label{figure-DL}
\end{figure*}

Research has shown that large language models (LLMs) possess exceptional capabilities in understanding complex contexts, answering questions, and generating content. In recent years, there has been growing interest in applying LLMs to various tasks in the financial domain. These applications are not only reshaping the landscape of financial analysis but also providing new perspectives for understanding market behavior and economic activity. Figure \ref{figure-DL} illustrates the workflow of LLM-based enterprise risk research. In the following, this paper provides a detailed discussion from three aspects: Few-shot/Zero-shot Learning, fine-tuning, and pre-training.
\\
\textbf{1. Few-shot/Zero-shot Learning}
\begin{figure*}[t]
    \centering
    \includegraphics[width=0.92\textwidth]{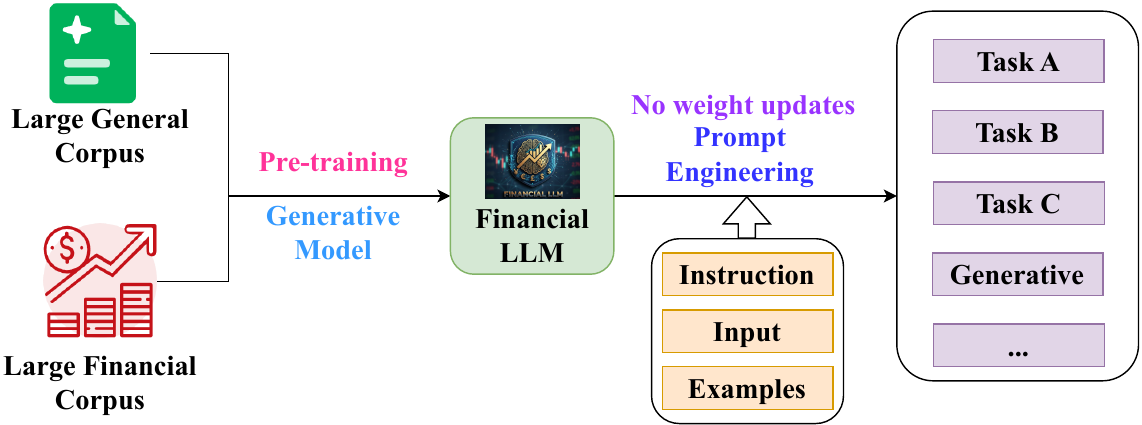}
    \caption{The workflow of LLMs prompt engineering.}
    \label{figure-ZF}
\end{figure*}
LLMs can quickly adapt to and perform entirely new downstream tasks that they have never been trained on, using only natural language instructions or a few contextual examples. This greatly reduces reliance on massive labeled datasets, overcomes the problem of data sparsity, and avoids the costly and time-consuming fine-tuning of LLMs, making it crucial for enhancing model generalization and enabling rapid deployment. The overall framework is depicted in Figure \ref{figure-ZF}.

Some scholars have attempted to directly leverage LLMs for enterprise risk-related research. For example, Fu \textit{et al.} \cite{fu2024exploration} indicate that incorporating advanced language models into the investment decision-making process can improve prediction accuracy and enhance the effectiveness of quantitative trading strategies. Similarly, Chen \textit{et al.} \cite{chen2023chatgpt} utilize ChatGPT’s graph inference capabilities to extract dynamic network structures from textual data for subsequent stock price movement prediction tasks. 
\\
\textbf{2. Pre-training}
\begin{figure*}[t]
    \centering
    \includegraphics[width=0.90\textwidth]{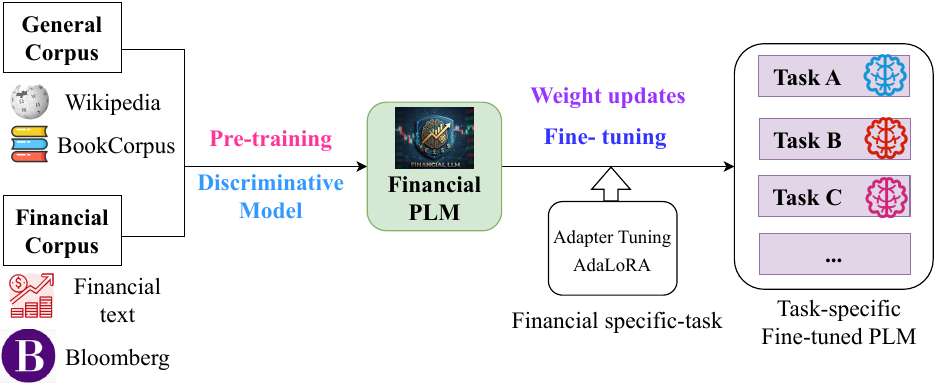}
    \caption{The workflow of LLMs pre-training and fine-tuning.}
    \label{figure-pf}
\end{figure*}
The aim of training LLMs from scratch is to create models that are more finely tuned to the finance domain. Table \ref{fintune-llm} lists the current finance LLMs that have been developed from scratch, including BloombergGPT, Xuan Yuan 2.0 \cite{zhang2023xuanyuan}, and Fin-T5. As illustrated in Table \ref{pretrain-llm}, there is a noticeable trend toward integrating public datasets with finance-specific datasets during the pretraining phase. Notably, BloombergGPT exemplifies a model trained on a corpus that consists of an equal mix of general text and finance-related content. It is important to highlight that BloombergGPT primarily utilizes a subset of 5 billion tokens specifically related to Bloomberg, which constitutes only 0.7\% of the total training corpus. This focused corpus significantly contributes to the performance enhancements observed in finance benchmarks.\\

\textbf{3. Fine-tuning}

Based on pre-trained LLMs, further training is conducted using small, high-quality datasets specific to a domain or task, in order to optimize the model’s performance in vertical scenarios. Common techniques for fine-tuning LLMs generally fall into two main categories \cite{hu2022lora}: full Fine Tuning and parameter-Efficient Fine Tuning. Fine-tuning LLMs within the finance domain can significantly enhance their comprehension of domain-specific language and contextual nuances, leading to improved performance in finance-related tasks and the generation of more accurate and tailored outputs. The overall framework is depicted in Figure \ref{figure-pf}.

FinGPT \cite{yang2023fingpt} introduces an end-to-end framework for training and applying Financial LLMs within the finance industry. It employs the lightweight Low-rank Adaptation (LoRA) technique \cite{hu2022lora} to fine-tune open-source LLMs, such as LLaMA and ChatGLM, using approximately 50,000 samples. However, the evaluation of FinGPT is confined to finance classification tasks. In contrast, Instruct-FinGPT \cite{zhang2023instruct} fine-tunes LLaMA using 10,000 instruction samples sourced from two Financial Sentiment Analysis datasets, but it too exclusively assesses performance on finance classification tasks. Pavlyshenko \cite{pavlyshenko2023financial} used a PEFT/LoRA-based approach for fine-tuning the Llama2 to enable multitask analysis of financial news. 
% Yin \textit{et al.} \cite{yin2023finpt} propose FinPT, a novel approach for converting tabular financial data into customer profiles, which are then used for financial risk prediction on FinBench after Profile Tuning on large foundation models. 
% Koa \textit{et al.} \cite{koa2024learning} introduce the Summarize-Explain-Predict (SEP) framework, which employs a self-reflective agent and Proximal Policy Optimization (PPO) to enable LLMs to autonomously generate explainable stock predictions. 
PloutosGPT \cite{tong2024ploutos} implements a rearview-mirror prompting mechanism to guide GPT-4 in producing rationales, while a dynamic token weighting mechanism fine-tunes the model by identifying and emphasizing key tokens within these rationales. In addition to instruction tuning, LoRA and quantized LLMs \cite{ma2024era} have also been applied for more efficient adaptation in financial tasks, as seen in FinGPT, FinGPT-HPC, and Llama-based models \cite{pavlyshenko2023financial}. 
% Pei \textit{et al.} \cite{pei2024modeling} developed a framework for automatically extracting company risk factors from news articles. By fine-tuning Llama-2, they achieved improved performance on the risk factor identification task.
\begin{table*}[t] % [!ht]表格在文本中放置的位置参数（努力放在当前位置，实在放不下，将放在下一页的顶部）
\centering % 表格整体居中
    \caption{The overview of finetuned finance LLM.}
    \label{fintune-llm}
    \newcommand{\tabincell}[2]{\begin{tabular}{@{}#1@{}}#2\end{tabular}}
    \resizebox{0.86\textwidth}{!}{
    \begin{tabular}{c|c|c|c} 
    \toprule[1.1pt]
     \textbf{Model name}&\textbf{Finetune data size} & \textbf{Training budget} & \textbf{Model architecture}\\
    \midrule
        FinMA-7B&Raw: 70k, Instruction: 136k&8 A100 40GB GPUs&LLaMA-7B\\
    FinMA-30B&Raw: 70k, Instruction: 136k&128 A100 40GB GPUs&LLaMA-30B \\  
   Fin-GPT(V1/V2/V3)&50K&$<$ \$300 per training&ChatGLM, LLaMA \\ 
  Instruct-FinGPT&10K Instruction&8 A100 40GB GPUs, $\backsim$1 hr&LLaMA-7B\\ 
Fin-LLaMA\cite{todt2023fin}&16.9K Instruction&NA&LLaMA-33B\\ 
   Cornucopia (Chinese)&12M instruction&NA&LLaMA-7B\\ 
  \bottomrule[1.1pt]
\end{tabular}
}
\end{table*}
\begin{table*}[t] % [!ht]表格在文本中放置的位置参数（努力放在当前位置，实在放不下，将放在下一页的顶部）

\centering % 表格整体居中
    \caption{The overview of from scratch trained Finance LLMs.}
    \label{pretrain-llm}
    \newcommand{\tabincell}[2]{\begin{tabular}{@{}#1@{}}#2\end{tabular}}
    \resizebox{0.9\textwidth}{!}{
    \begin{tabular}{c|c|c|c} 
    \toprule[1.1pt]
     \textbf{Pretrained
LLM}&\textbf{Corpus size} & \textbf{Training budget (A100·hours)} & \textbf{Model architecture}\\
    \midrule
       BloomBergGPT&363B Finance tokens + 345B
public tokens&1,300,000 &50B-BLOOM\\
    XuanYuan2.0&366B for pre-training + 13B for
finetuning&Not released &176B-BLOOM \\  
   Fin-T5&80B Finance tokens& Days/weeks&770M-T5 \\ 
  \bottomrule[1.1pt]
\end{tabular}
}
\end{table*}

\section{Spotlights of Representative Works}
\label{section-spotlights}
%In this section, we first make a brief introduction to the rep-representative works and then present their unique contributions.
In this section, we begin with a brief overview of the representative works, followed by a discussion of their distinct contributions.

We select the representative works based on citations
and methodological progress. The citation for each reference is sourced from Google Scholar, with statistics current as of November 23, 2025. The highly cited studies can be categorized into two groups according to their focus: existence and methods. Table \ref{table-Contributions-Works} summarizes the key findings of these representative works for convenient reference.

The first category is the  existence category, which focuses on demonstrating the comovements between enterprise financial risk and various data types, including financial indices, textual information, relational data, and integrated intelligence. For instance, Belas \textit{et al.} \cite{Daily1994Bankruptcy} conduct a study revealing a direct link between corporate governance structures and enterprise bankruptcy. Imbierowicz \textit{et al.} \cite{imbierowicz2014relationship} explore the relationship between liquidity risk and credit risk in the banking sector from multiple perspectives. Additionally, Chae \cite{chae2015insights} analyze supply chain-related tweets, highlighting Twitter's potential impact on supply chain practices and research.

The second category comprises methodological studies, which focus on leveraging various techniques—especially recent advances in deep learning and LLMs—to extract valuable information from textual media and relational data. This approach helps establish connections between multi-source heterogeneous data and corporate financial risk. For example, FINCON \cite{yu2024fincon} adopts a manager–analyst hierarchical structure, enabling multiple functionally distinct agents to collaborate synchronously through natural language interactions, and it achieves strong performance in both single-stock trading and portfolio management tasks.

\begin{table*}[htb]
    \centering
         \caption{Contributions of Representative Works.}
         \label{table-Contributions-Works}
         \newcommand{\tabincell}[2]{\begin{tabular}{@{}#1@{}}#2\end{tabular}}
        \resizebox{\textwidth}{!}{
           \begin{tabular}{l|c|l}
            \toprule[1.1pt]
            \textbf{Group} &   \textbf{Reference} & \textbf{Contribution}  \\
            \midrule
            %  &\cite{Altman1968Financial}& This work first \\
             
            %  &\cite{Acemoglu2015Systemic} & \\
            %  &\cite{Lubatkin2006Ambidexterity} &  the management capabilities\\
             
            %  & \cite{Rosenbusch2011Is} & the innovation ability  \\
            %  & \cite{Altman1994CorporateDD} & using linear discriminant analysis and neural networks \\
            %  \\
            %  \midrule
             % Existence&\cite{daily1994bankruptcy}&\tabincell{l}{The authors used logistic regression to analyze the relationship between bankruptcy and corporate governance structure, \\which gave rise to the trend of studying the relationship between corporate governance structure and financial position.}\\
            \multirow{6}{*}{Existence}
            % &\cite{cachon1999competitive}&\tabincell{l}{The authors use a game model to study the relationship between inventory strategy and competitive cooperation in \\supply chains, and the conclusion that competition may reduce inventory brings a new insight to the study of supply chain risk.}\\
            % &\cite{beasley2005enterprise}&\tabincell{l}{The authors examine what are the reasons for companies to adopt ERM and the conclusions provide a preliminary basis \\for future research on ERM deployment.}\\
            &\cite{imbierowicz2014relationship}&\tabincell{l}{This is the first paper to examine the relationship between liquidity risk and credit risk in the banking industry\\ from different perspectives, and the findings have significant implications for subsequent research. } \\
              
              % &\cite{dichev1998risk}
              % &\tabincell{l}{\tabincell{l}{Using Z-score and O-score model for bankruptcy prediction, it demonstrates that higher bankruptcy risk doesn’t \\ bring about higher returns which is different from previous opinions.}}\\
              
              % &\cite{ericsson2009determinants}
              %  &\tabincell{l}{\tabincell{l}{It analyzes the linear relationship of theoretical determinants of default risk and default swap spreads and identifies\\ explanatory power of volatility and leverage in regressions. }}\\
              %   &\cite{he2012rollover}
              % &\tabincell{l}{\tabincell{l}{This paper puts forward a model to analyze the influence of bond liquidity on enterprise credit risk, which provides\\ a new idea for enterprise credit risk management.
              % }}\\
              
             %&\cite{maghyereh2021effect}&\tabincell{l}{This study fills a key gap in the impact of oil market shocks on systemic banking risk and extends the relevant literature by examining\\ both the global financial crisis and the COVID-19 pandemic.} \\ 
             &\cite{tang2008power}&\tabincell{l}{The paper proposes a unified framework. The findings highlight the power of flexibility while clarifying the associated\\ benefits of different levels of flexibility, providing insights into deploying flexibility to reduce supply chain risk. }\\
            &\cite{craighead2007severity}&\tabincell{l}{This article examines the impact of supply chain disruptions. It adds to existing knowledge and also questions the \\wisdom of pursuing practices such as reducing the supply base, global sourcing and sourcing from supply clusters.}\\
                   
            \midrule
            \multirow{10}{*}{Methods}
            &\cite{billio2012econometric}&\tabincell{l}{The authors use principal component analysis and Granger causality networks to study systemic risk among banks, \\insurance, funds and dealers, the method has been influential in studying systemic risk.}\\
            
            % &\cite{barboza2017machine}
            %   &\tabincell{l}{\tabincell{l}{This study tests the performance of machine learning models in bankruptcy prediction and it finds a higher average \\performance than usual models. Thus, it’s a substantial improvement in accuracy using machine learning methods.}}\\
            % &\cite{maloni2006corporate}&\tabincell{l}{The framework of CSR in food supply chain proposed in this paper lays a foundation for the food industry to further \\study the elements of CSR in supply chain.} \\
            &\cite{kou2014evaluation}&\tabincell{l}{In this paper, an opportunistic MCDM clustering algorithm is proposed to rank the popular algorithms in the field of \\financial risk analysis. Finally, the MCDM method is proved to be the most effective in evaluating clustering algorithms.} \\

        &\cite{yang2021Financial}&\tabincell{l}{This paper is the first to employ graph neural networks and proposes a novel financial risk analysis framework based \\on supply chain mining among enterprises.}\\
        &\cite{cheng2020contagious}&\tabincell{l}{This paper is based on deep neural networks and employs a temporal inter-chain attention network to model loan \\behavior data within guarantee networks.}\\
        &\cite{araci2019finbert}&A BERT-based language model designed for NLP tasks in the financial domain.\\
        &\cite{wu2023Bloomberggpt}&\tabincell{l}{BloombergGPT is a 50-billion-parameter language model trained on the largest domain-specific dataset to date.}\\
            \bottomrule[1.1pt]
           \end{tabular}
          }
        %  }
    % \end{center}
\end{table*}

%%%%%%%%%%%%%%%%%%%%%%%%%%%%%%%%%%%%%%%%%%%%%%%%%%%%%%%%%
\section{Directions For Future Work}
\label{section-futurework}
\subsection{Enterprise Intelligence}
We anticipate that enterprise risk prediction will increasingly rely on the integration and analysis of multi-source data. These data sources may include information from customers, industries, affiliated companies, and government agencies, while seemingly insignificant data—such as corporate utility bills—can also serve as valuable supplements to traditional datasets. The complexity and high-dimensional nature of financial data introduce unique challenges, and data pollution may become a multifaceted issue that can significantly affect the performance and reliability of LLM-based models. Moreover, it is important to recognize that data availability remains a major challenge in enterprise risk prediction. Therefore, prioritizing the development of open databases is essential, as they can serve as platforms for data sharing. By facilitating access to a broader range of data, we can unlock its full potential and achieve a more comprehensive understanding of enterprise risk profiles.
\subsection{Analysis Model}
Balancing the need for fast and cost-effective model inference with performance requirements is a significant challenge, as LLMs typically incur substantial computational overhead. This can result in high inference costs and slower processing speeds, particularly when dealing with large datasets. When applying LLMs to generate content for financial tasks, legality and reliability become critical concerns. Financial reports are subject to strict legal and regulatory standards, and any inaccuracies may lead to severe consequences. Moreover, because LLM outputs are sampled from a distribution rather than being deterministic, estimating uncertainty and providing confidence intervals for model predictions is especially important in the financial domain.
\subsection{Contagion Mechanism}
Current research is focused on the potential risks that financial networks present to companies. To fully understand these risks, it is essential to investigate the specific pathways through which they are transmitted. This includes assessing the impact of risk on individual companies and examining the extent of risk propagation across the network. Additionally, it is important to explore the temporal dynamics of risk transmission, such as the duration required for risks to spread to related companies. Another significant aspect is whether the types of risks faced by related companies align with those of the original company. Unfortunately, technological limitations have hindered comprehensive studies on these critical questions. In light of globalization, it is imperative to develop a thorough understanding of the mechanisms of risk transmission. Thus, further exploration in this research direction is highly recommended.

\subsection{Risk Interpretability}
Although LLMs have achieved some success in enterprise risk analysis, they often operate as black boxes, failing to explain the rationale behind their predictions. This opacity poses challenges for their application in critical domains such as finance and security. Traditional interpretability methods can be broadly categorized into two types: instance-level methods and model-level methods, which are often not directly applicable in real-world financial scenarios. Therefore, exploring a risk detection framework that can simultaneously provide high-quality predictions and meaningful explanations represents a promising direction for future research in enterprise financial risk analysis. Additionally, developing standardized metrics to evaluate the quality of generated explanations can help stakeholders better understand and effectively leverage AI-generated insights.

\subsection{ Ethical Issues}
Given the significant threats posed by data breaches and compliance violations, the security and privacy of financial data are of paramount importance. Deploying LLMs in the financial sector presents substantial challenges in maintaining robust data protection measures and safeguarding sensitive information. Implementing strong security protocols and advanced cybersecurity techniques can mitigate the risk of data leaks and ensure compliance with privacy regulations, thereby building trust and protecting sensitive data. To further prevent data leakage, techniques such as federated learning and local deployment of LLMs can be adopted, enabling organizations to ensure the security and privacy of financial data while still benefiting from the advanced capabilities these models provide.
% LLMs have a wide range of applications across various sectors, including intelligent customer service, investment advisory, marketing, risk management, operations, investment research, investment banking, and quantitative trading. In the financial industry, organizations like Morgan Stanley, Stripe, Bloomberg, and FinGPT have already begun to leverage LLMs \cite{wu2023bloomberggpt, yang2023fingpt}. Within the realm of enterprise financial risk, LLMs can play a crucial role in public opinion analysis and the extraction of regulatory information from textual sources. Additionally, LLMs can enhance existing financial NLP tasks, such as relation extraction and named entity recognition, facilitating the construction of comprehensive enterprise knowledge graphs. Subsequently, the inference capabilities of LLMs on graph-structured data can be harnessed to conduct in-depth analyses of financial risks faced by enterprises.

% \subsection{Risk Management}

% Figure \ref{figure-interpretablity} shows ...
% \begin{figure}[htb]
%     \centering
%     \includegraphics[width=0.3\textwidth]{interpretablity.pdf}
%     \caption{Interpretability.} 
%     \label{figure-interpretablity}
% \end{figure}

%%%%%%%%%%%%%%%%%%%%%%%%%%%%%%%%%%%%%%%%%%%%%%%%%%%%%%%%%
\section{Conclusion}
\label{section-conclusion}
In this survey, we attempt to cascade and systematize existing research on enterprise financial risk, providing a comprehensive review of enterprise risk analysis. This is the first survey for enterprise financial risk from the the perspectives of big data and large language models. In particular, we first begin with an introduction to the types, granularity, intelligence and evaluation metrics of enterprise financial risk. Then we classify models of enterprise financial risk, show the basic implementation framework for each type of model, and summarise representative approaches in enterprise risk analysis. Finally, we elaborate on current cutting-edge research and its possible future directions. We believe that this survey will provide researchers in the field with the fundamental knowledge and a clear framework for research.

\subsubsection{\ackname} We would like to express our gratitude to Binchen Yang, Xuesong Zhang, Meiai Wang, Xiaoying Sun, Qiaoyue Liang, Yiying Yuan, Lanqi Zhao, Yuansong Zhu, and Ying Liu for their support in literature collection.

This research is partially supported by funding from Xiangjiang Laboratory \\(25XJ02002), the National Natural Science Foundation of China (62376228, 62376227), the Science and Technology Innovation Program of Hunan Province (2024RC4008), the China Postdoctoral Science Foundation (2025M770766), Sichuan Provincial Postdoctoral Research Project Special Funding (TB2025043) and Sichuan Science and Technology Program (	2026NSFSC1496, 2023NSFSC0032). Carl Yang is not supported by any funds from China.
% \section*{Acknowledgments}
% The research is supported by the National Natural Science Foundation of China under Grant Nos. U1811462, 71725001, 71910107002, 62376227, 61906159, 62176014, 71873108, 62072379, and Sichuan Science and Technology Program under Grant No. 2023NSFSC0032, 2023NSFSC0114, and Guanghua Talent Project of Southwestern University of Finance and Economics, and Financial Innovation Center, SWUFE (Project NO.FIC2022C0008) and ``Double-First Class" International Innovation Project (SYL22GJCX08), and Fundamental Research Funds for the Central Universities (JBK2304150).


\end{document}